# A Human Dimension of Hacking: Social Engineering through Social Media

**Heidi Wilcox and Maumita Bhattacharya***

School of Computing & Mathematics, Charles Sturt University, Australia

*Corresponding author

**Abstract.** Social engineering through social media channels targeting organizational employees is emerging as one of the most challenging information security threats. Social engineering defies traditional security efforts due to the method of attack relying on human naiveté or error. The vast amount of information now made available to social engineers through online social networks is facilitating methods of attack which rely on some form of human error to enable infiltration into company networks. While, paramount to organisational information security objectives is the introduction of relevant comprehensive policy and guideline, perspectives and practices vary from global region to region. This paper identifies such regional variations and then presents a detailed investigation on information security outlooks and practices, surrounding social media, in Australian organisations (both *public* and *private*). Results detected disparate views and practices, suggesting further work is needed to achieve effective protection against security threats arsing due to social media adoption.
**Keywords:** Social engineering; Social media; Human dimension of hacking; Information security.

## 1. Introduction

Social engineering is one of the crucial human dimensions of hacking and social media is emerging as the easy enabler. The distinction between personal and organizational social circles is becoming increasingly blurred with the introduction of social media in various forms throughout the organization. Traditional security countermeasures are not keeping up as more businesses are encountering breaches targeting the human elements, such as social engineering. Most employees possess a natural inclination to trust others and disclose information in a well-meaning, helpful manner. Social engineers exploit these human idiosyncrasies in behaviour to form an attack from the outside that leads them to gain inconspicuous entry into protected areas of the company for their own illicit use [10, 11, 12, 14]. The very nature of social networking encourages people to trust and engage with their communities; however, this trust can very easily be exploited by cyber-criminals whose intention is to prey on victims. A holistic framework for security should combine technical, procedural and user-centric controls along with process guidance; and is recommended for the ever-evolving technologies embraced by organizations, such as social media adoption [3, 8]. Social media policies and guidelines provide necessary advice on how social media participation will be applied to all of the members of an organization. Current efforts by organizations in social media policy development primarily see inclusions for legal clauses, disclaimers; and guidelines for participant online conduct or content creation. However, perspectives and practices regarding organisational information security issues surrounding social media, vary from region to region. In this research, we explore such perspectives and practices of global regions, including a detailed study based on Australia.

The rest of the paper is organised as follows: Section 2 explains the concept of enterprise social engineering, while Section 3 outlines relevant perspectives and practices in various regions of the world. In Section 4, we present some findings from our Australia-based study. Finally, concluding remarks are presented in Section 5.

## 2. Enterprise Social Engineering

Social engineering is the art of manipulating people into performing a specific task for the engineer rather than the engineer gaining entry into systems or networks using the traditional technical hacks [10, 11, 12]. The approach that social engineers employ when initiating and completing these attacks can be represented by a lifecycle depicting certain patterns that have been interpreted by a number of researchers in the area of social engineering. Zulkurnain, Hamidy, Husain and Chizari in [15] provide phase-based models similar to our adaptation of the Social Engineering Lifecycle as introduced in [12] and also briefly explained below.

In [12] we introduced a Social Engineering Lifecycle including the following four phases: (a) Fact-finding – The phase to gather information about the target and then use this information to build a relationship with the target or someone relevant to the success of the attack. (b) Entrustment – In this phase, the aggressor tries to position himself into a position of trust. (c) Manipulation – The target is manipulated by the "trusted" aggressor into revealing information, such as, password or perform an action to the benefit of the aggressor. (d) Execution: The cycle is completed as the target/victim completes the task or tasks requested by the aggressor.

The main individual data breach sources associated with social engineering are phishing attempts from social media sites, emails or insecure mobile devices. However, the combination of these sources provides a very different overall representation. Individuals are increasingly accessing social networking sites from mobile applications on smartphones.

Social engineers are now predominantly using social media to stalk and target victims, using the information gathered to lure their victim with phishing techniques – applied through the site itself or by traditional email [10, 12]. Additionally, the inherent insecurities of mobile technologies are monumentally contributing to the success and covertness of these attacks.

## 3. Social Media & Information Security: Regional Perspectives

This section presents an outline of social media and information security scenarios from various global regions, based on literature review.

### 3.1. North America

With the inclusion of the United States, this area is considered a leader in cyber-maturity and cybersecurity awareness [6]; the public–private partnership is rather nebulous though [1]. It is noted that the US government is highly targeted for advanced persistent threats from various nations and has more than likely been forced into advancing online protection for the homeland's defense and critical infrastructure. The US government encouraged social media adoption into many government process, and has subsequently been proactive in seeking up to date knowledge and security regarding these technologies. These security efforts have seen an increase in publications for the general public to increase awareness and best practice within their own online communities. Some older studies [13], found that US government divisions are very outspoken compared to countries, such as Korea on matters of citizen privacy and security in online social media use; however, these concerns are not always followed through with adequate governance and policy. Alternatively, industry-based research finds that organizations within the United States are more likely to have an online security strategy and include policies for social media and personal device usage in the workplace, compared to other nations worldwide [6, 9].

### 3.2. South America

Brazil and Mexico are seen as rapidly developing nations within internet and mobile penetration user statistics [9]. Brazil, with the hosting of soccer's 2014 World Cup, saw an explosion of user device and social media to share excitement over the highly revered global sporting event. This is an example that demonstrates, cultural events, such as the World Cup, can also be an enabler to technological adoption and development, consequently leading to strategy for risk management of these technologies. This is reflected in studies showing that Brazil was one of the most improved and leading nations in creating increased focus to information security budgets and user focus [9].

*3.3. Middle East/Africa*
Despite many African countries initially incurring relatively poor internet infrastructure and online capacity problems with an urban-rural gap in connectivity [4] and Middle Eastern Arab perimeters embargoing social media attempts by users, these nations are rapidly overcoming these barriers, using the population levels within African countries to combine with the wealth of the Gulf States, to increase ownership of mobile device and social media penetration.

*3.4. Europe*
This continent holds somewhat fragmented views of how cybersecurity should be approached [2]. The EU Commission and the European Union Agency for Cybersecurity (ENISA) urge European nations to combine forces in combatting cybersecurity. Academic evidence of this can be found in research by Fast, Sorensen, Brand & Suggs [5], which examines the existence of policy for social media in European public health organizations. This study found 75% of the 21 organizations chosen had a social media policy; with the major coverage areas relating to data and privacy protection, intellectual property and copyright protection and social media participation regulation. Additionally, the UK as a global country is less likely to show concern for information security and data breaches which are ever-increasing due to high levels of social media and mobile penetration. Industry-based studies show that compared to the worrying amount of data breaches attributed to human error through social engineering and phishing, user training and awareness to these issues remain a low priority [9].

*3.5. Australia & Asia-Pacific*
Australia and many Asian neighbours are viewed as technologically advanced nations, especially due to research and development (R & D) capacities and facilities located in India, China and Japan [6]. Despite Oceania and Asia-Pacific countries displaying various regional cultures, demographics and social models, populations are either already high adopters of online participation (China, Australia, South Korea, Japan); or noted as rapidly emerging technological markets (India, Indonesia, Thailand) [6].
The Australian government is leading efforts in encouraging users to adopt social media for online participation and engagement within communities and is increasing efforts at awareness of immediate threats aimed at business and consumers. The Australian Cyber Security Centre (ACSC), hosted by the Australian Signals Directorate, was formed to include experts in their fields representing the Defense Intelligence Organization, Australian Security Intelligence Organization, the Attorney General's Department, Australian Federal Police, Australian Crime Commission and so on. The ACSC analyses and assesses incoming cyber threats, and works closely with various industries and private sector parties to formulate effective countermeasures and create public awareness. The government also published guidelines and policies, primarily directed at government departments; however, they are publically available to all organizations for best practice guidance.

*3.6. Overall Trend*
Overall, trends indicate that corporations are increasingly recognizing online threats as an overall business or strategic risk as opposed to being a technology risk. Figure 1 highlights our summation of information security strategies via a broad continental perspective. This view amalgamates key concepts in the data obtained through literature review.
Overwhelmingly, all industries are discovering the need to establish trust and enhance business process transparency due to the pervasive and instant nature of social media [7]. Reputations can be lost or gained rapidly through the voice of the online communities.

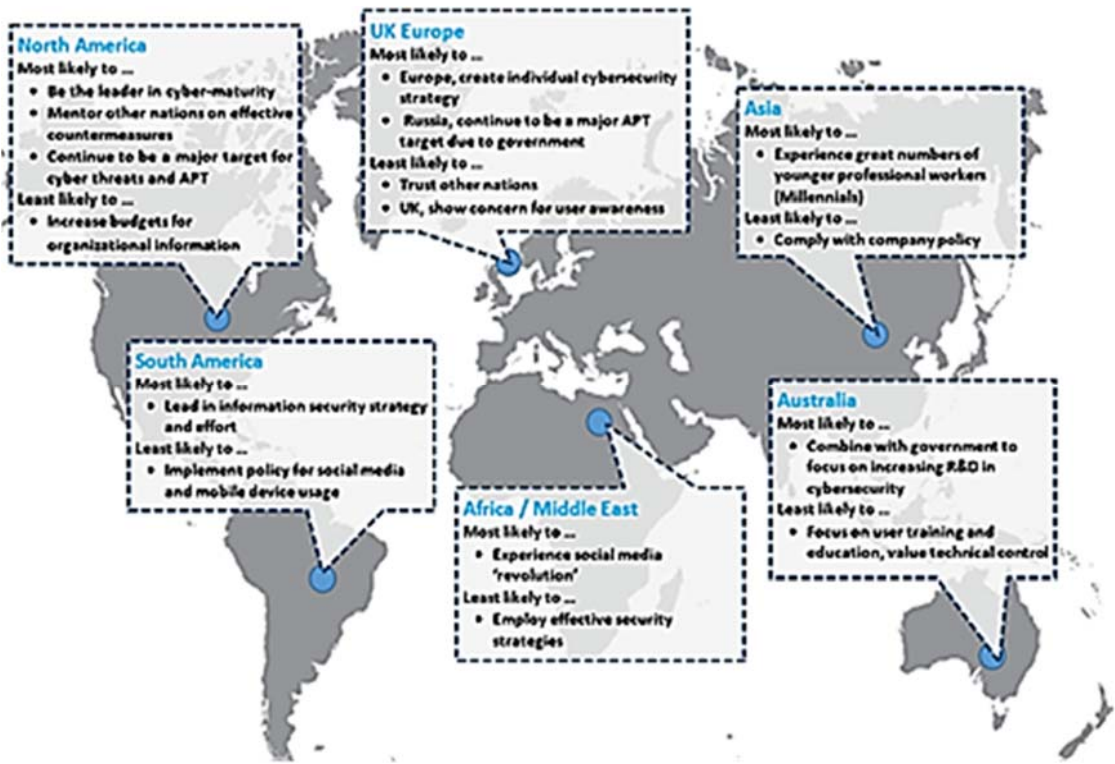

**Figure 1.** Representation of global online environments and cybersecurity concerns.

The increased adoption of employees using social networking for both business benefit and personal socializing has created an immediate need for management to re-assess their current security culture, making employees a primary focus. This however, is happening at a slower pace than these technologies are being introduced and also the traditional policies and training are not effectively covering all aspects of information security. The results are that there are more cyber-attacks related to online social engineering than ever before.

**4. An Australian Study**

In this section we present the result of a descriptive survey-based study, we conducted, to measure the state of employee related issues faced by social engineering through social media, within Australian public and private sector organizations.

*4.1. Methodology*

We used a mixed approach to study the perspectives and practices of social engineering through social media within Australian organizations, including both public and private sectors. The questionnaire was developed, piloted and then administered by invitation to Australian senior managers and policy developers, information security personnel and end users who may, or may not, use social media as part of their daily activities within organizations. The survey included both closed and open ended questions, broadly categorized in areas relating to (i) people, (ii) process and (iii) technology to gather data on social media management (including policy issues), awareness of social engineering and technical threat mitigation measures.

Due to the categorization of potential survey participants, we invited 80 Managers and policy developers, 80 information security personnel and 200 end users to respond to the aligned question set within the questionnaire.

As mentioned above, the original study investigated three aspects, namely, people, process and technology. However, in this paper, we shall report only the process component. The process component of the questionnaire dealt with existing business process management techniques, such as risk management, social media strategy and governance issues. This section of the questionnaire was applicable to all research participants through some guided questions; however, due to its management focus, a more detailed question set was aimed at senior managers and policy developers only.

*4.2. Results and Discussions*

Within the 'process' component, our original study looked at the following three aspects: (i) Social media management, (ii) Business risk and consequences and (iii) Policy and governance issues. In this paper, we are reporting just the first two of the three aspects mentioned.

*4.2.1. Social media management.* In an interesting selection of answers in this section, 70% of management personnel stated that their organization distinctly guides or promotes a separation of personal and professional social media use. However, half of the end users that responded to the same question admit they were not completely sure of these distinctions, leading to the worrying conclusion that employees are confused of established boundaries that must exist between their use (especially comments and posts) for personal opinion and views as opposed to the official stance of the company. Not provisioning this clear distinction to end users opens up the organization to a multitude of legal and reputational damage issues and could enhance the manipulative techniques of social engineers when baiting or phishing its users.

*4.2.2. Business risk and consequence.* Concerning risk management of social media technologies, the majority of management within the public or private sector (regardless of industry) feel very concerned about the associated issues and threats posed to crucial assets. A small number also believe that the adoption of these technologies are worth the risks, with 20% of all managers confidently consider that these risks can be avoided or mitigated. From a manager's perspective, concern is very high for social media risk having negative impact on protection of customer data or personally identifiable information (PII) (67%) and information leakage (50%). Other organizational social media risk that rated with high importance included company fraud, incorrect information, increase in malware/virus, reputation damage and employee bandwidth usage. These consequences with a rating of very high or high are provided with more detail in Figure 2.

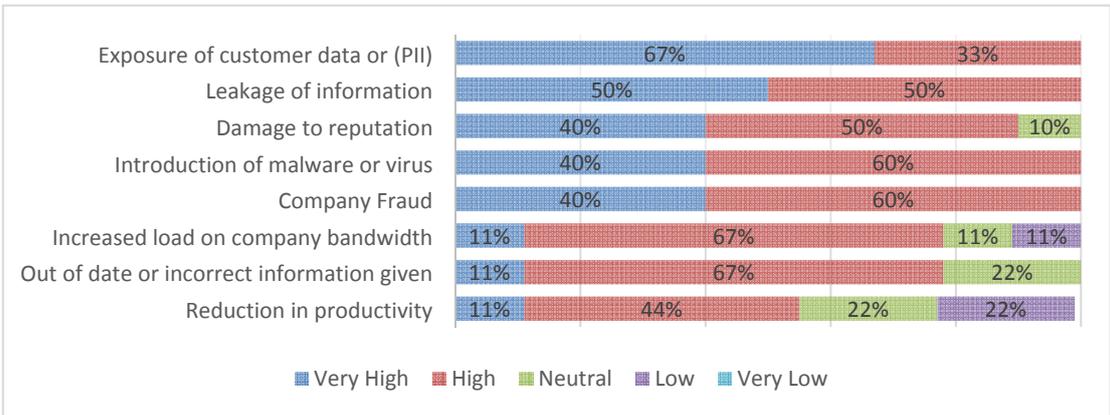

**Figure 2.** Management rating of risk consequence owing to social media threats

**5. Conclusions**

In this paper we discoursed one of the evolving human dimensions of hacking, namely, social engineering through social media. A snapshot of region-based social media and relevant information security scenarios of the world was presented. We have also presented some results from our Australia-based study about perspectives and practices of organisations (both *public* and *private*), regarding security threats associated with social media usage; social engineering in particular. Further to what has been already reported in Section 4, as per our Australia-based study, the private sector managers were very concerned with social media risks and rated company fraud and reputation damage above all others. In contrast, public sector managers were aware of the risks; thought the benefits outweighed the risks and that these risks could be avoided or mitigated. Their concerns focused quite rightly on breaches of customer data and the prescription of incorrect or out-of-date information.